# A Novel Proposal for Manufacturing Steel: OSRAM's $CO_2$ Steel Making Process

O. Seetha Ramayya[1] and O.S.K.S. Sastri[2]

[1]Alumini, Indian Institute of Science, Bangalore and Technical Advisor Retd. (FACOR), Sriramnagar, A.P., India

[2]Department of Physics and Astronomical Sciences, Central University of Himachal Pradesh, Dharamshala, India

**Abstract**

*In this paper, we propose a concept to utilise Carbon-dioxide for Steel-Making based on some initial experiments and would like to name it after the first author as OSRAM-$CO_2$-SM process. We found in our lab experiments that the carbon content in high-carbon ferro-chrome metallic powders has come down from 7% to 1% when pure $CO_2$ gas with partial pressure of 1 atm is passed through a horizontal retort furnace maintained at $1100^o$ C for 24 hours. Our results clearly demonstrated that decarburisation can be very effective, when $CO_2$ is used at temperatures where it is unstable and the Boudouard reaction which is endothermic in nature is more favoured. Based on these findings, we propose the OSRAM-$CO_2$-SM process, a concept paper, in which $CO_2$ shall be used as the decarburising agent. $CO_2$ has to be passed into a specially designed converter provided with heating mechanism to maintain the contents from blast furnace in a molten state for decarburisation of the melt. The out coming hot gases consisting of CO has to be burnt in a combustion chamber with stochiometric proportions of pure O2 to produce CO2 at a partial pressure close to 1 atm. The process will be extremely useful as it involves reuse and recycle of $CO_2$ and in turn would reduce the overall amount of $CO_2$ discharged into the atmosphere.*

**Keywords**: Decarburisation, Greenhouse gases, Steel-Making

## 1. Introduction

About three billion tonnes of Carbondioxide is emitted and an almost equal amount of Oxygen is consumed in making about one hundred and seventy million tonnes of steel per annum [1]. This is one of the major contributers to greenhouse gases and is of great environmental concern. Decarburisation is a major step in steel making where the carbon content in the molten iron from the blast furnaces is brought down from about 4% to less than 1% by blowing very expensive pure oxygen into the Basic Oxygen Furnace (BOF) [2-4] which was an improvement over the process originally proposed by Bessemer. Alternative steel-making involves Electric Arc Furnaces (EAF) [4,5] where scrap, alloy scrap, molten metal and ferro alloys are used for producing alloy and stainless steel (with less than 0.05% carbon), by blowing oxygen along with other inert gases. But, predominantly it is used in secondary steel-making process. Recently, Molten Oxide Analysis has been suggested as an innovative alternative to steel-making with lower $CO_2$ emissions [6].

Studies [7-9] regarding decarburisation of Fe-Cr-C and Fe-C in both solid and liquid states have shown that $CO_2$ can be effectively used as an oxidising agent. We have also performed experiments with $CO_2$ as decarburising agent in the laboratory, which has motivated us to propose the current process. The lab experiment and the results are presented in the next section. Section 3 deals with the O. SeetaRAMayya (OSRAM) $CO_2$ Steel-Making Process in detail. In Section 4, we discuss the viability of the process with the emerging technologies and the merits of the proposed process over the existing conventional steel-making processes. Finally, we conclude in Section 6 with prospects for future work.

## 2. Experimental Results for Decarburisation of High Carbon Fe-Cr powder using $CO_2$:

The experimental setup for the decarburisation of high carbon ferro-chrome powder performed at Ferro Alloys CORporation Ltd (FACOR) is shown in Figure.1. 100 gms of metallic powder of Fe-Cr with the composition of 72% Cr, 7% C, 2% Si and 19% Fe has been placed in a ceramic boat inside the horizontal retort furnace which is maintained at a constant temperature of 1100°c. Pure $CO_2$ gas from cylinders is passed at 1 atm pressure for 24 hours through the furnace at a very slow rate of 2 $Nm^3$ per hour.

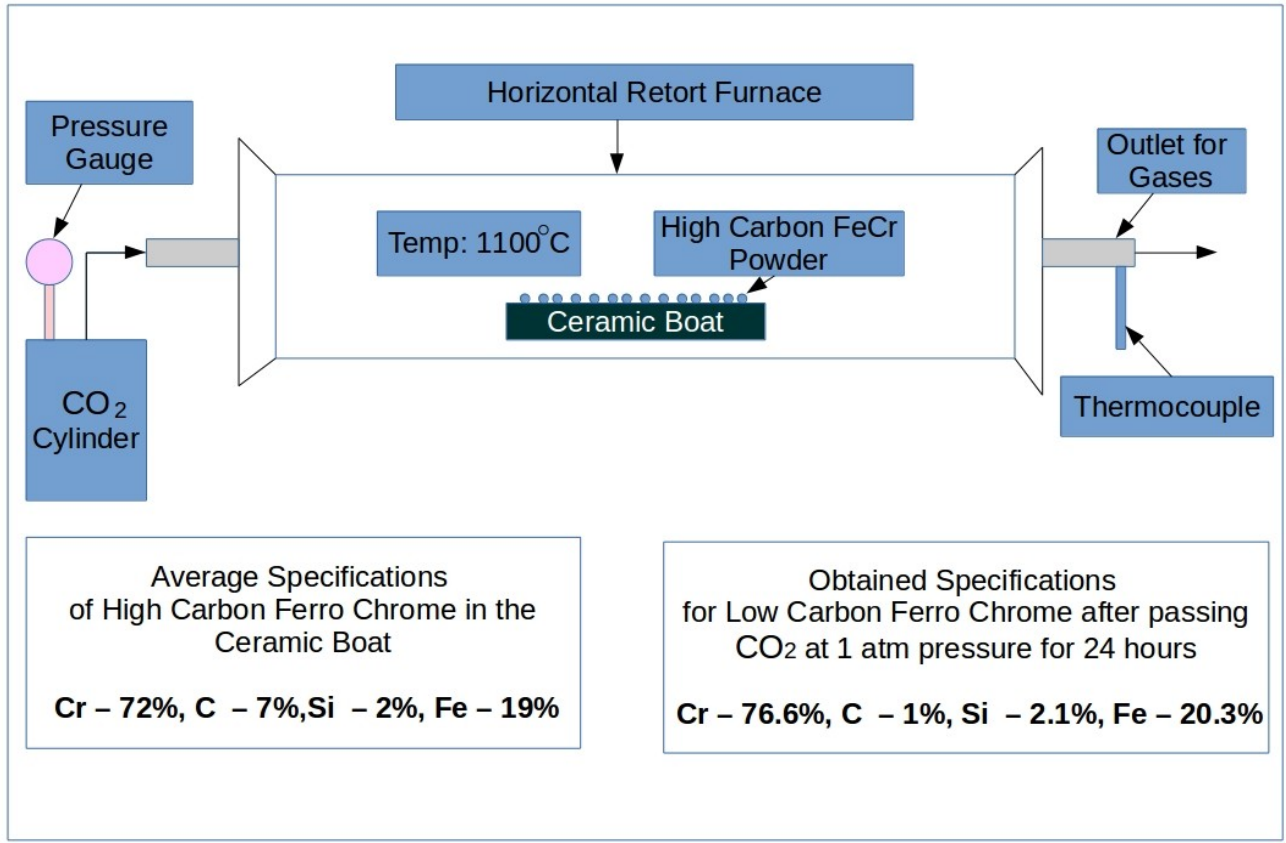

**Figure 1:** Scematic diagram of the experimental setup for decarburisation of HC Fe-Cr powders along with average specifications of feed and output.

The experiment was repeated 8 times and reproducibility of the results were ascertained. The final product consisted of 94 gms of the Low Carbon Fe-Cr powder which was chemically analysed and the percentage compositions are given in Figure.1. The results clearly indicate that carbon content has been brought down from 7% to 1% without loss of other elements. Wang Haijuan et.al.[9], have demonstrated that blowing $CO_2$ through high carbon Fe-Cr-C melts have proved beneficial in the sense that decarburisation takes place without any loss of Chromium till the carbon content in the melt reached 0.08%. Bhonde.P.J. and Angal. R.D [8] have shown that $CO_2$ can be utilised for solid state decarburisation of high carbon ferro-manganese very effectively. All these results have motivated this effort to propose a novel method of steel production named as the OSRAM-$CO_2$ steel-making process after the first author, which is discussed in the next section.

## 3. OSRAM-CO$_2$ Steel-Making Process:

The existing steel-making process uses very expensive and energy consuming pure oxygen along with other gases such as air+Argon+N$_2$+H$_2$+steam etc., to decarburise the iron melt tapped from the blast furnace to low carbon iron preferably consisting of less than 1% carbon. The output of this BOF is typically CO along with small proportions of N$_2$, H$_2$ etc which are let out from the top of the converter and allowed to burn with the air present in the atmosphere and thus releasing large quantities of CO$_2$ to the environment. The various reactions that take place with dissolved oxygen are all exothermic in nature and are as follows:

$$Fe + O = FeO$$

$$Mn + O = MnO$$

$$Si + 2O = SiO_2$$

$$[C] + O = CO$$

$$2P + 5O = P_2O_5$$

These oxidation reactions provide the energy needed to melt fluxes and scrap to the desired temperature of liquid steel, thus making BOF an autogenous process with no necessity for any external heat source.

The schematic diagram of OSRAM-CO$_2$ Steel-Making process integrated with the conventional BOF is shown in Figure 2.

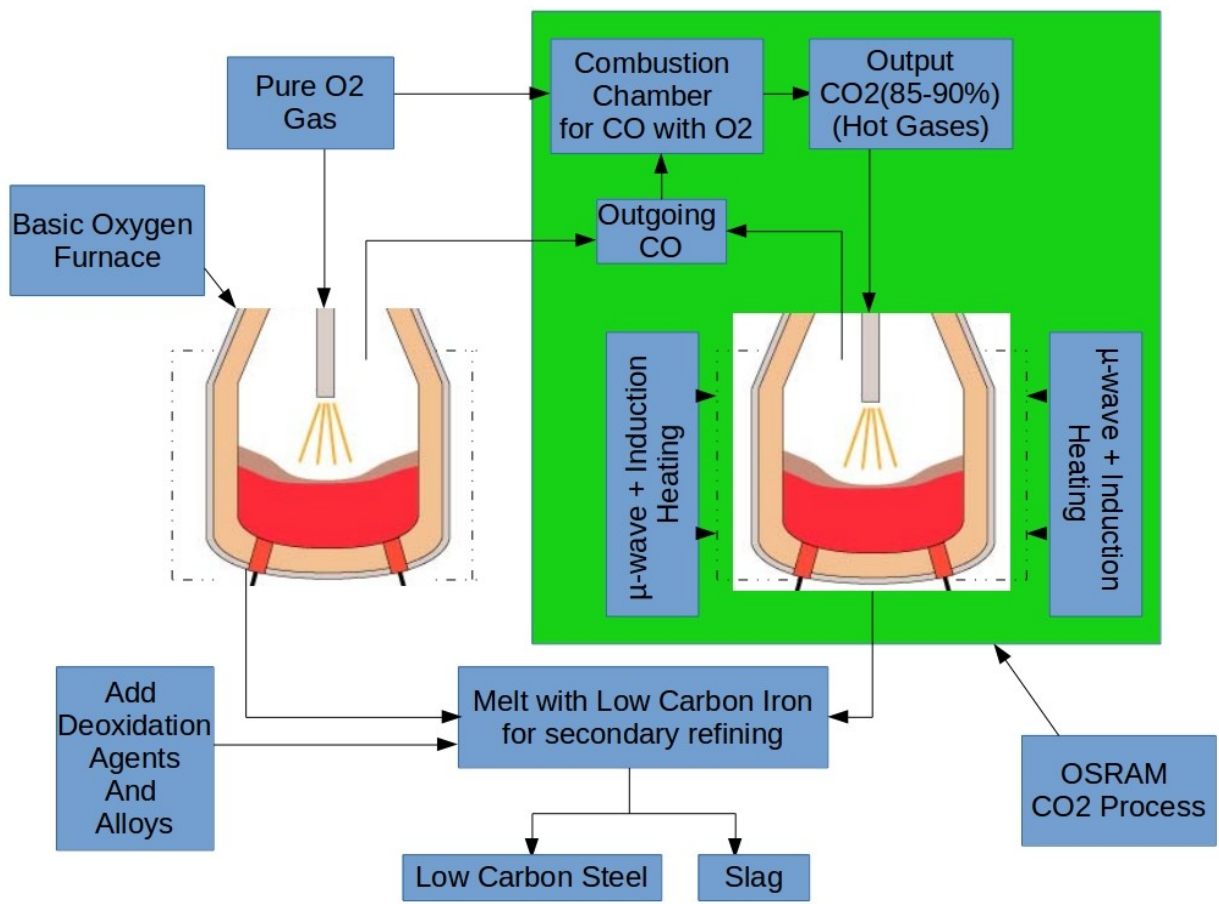

**Figure 2:** Schematic Diagram of OSRAM-CO2 Steel-Making Process (inside the green box). The converter to the left is the conventional BOF using O$_2$ and the one to the right is OSRAM converter with external heat source shown utilising hot CO$_2$ gases as decarburising agent. The out coming CO gases are recycled and burnt in the combustion chamber to produce CO$_2$ that can be resued in the OSRAM converter.

The fundamental idea in this process is to use $CO_2$ as a decarburising agent and reduce carbon content to less than 1% in the iron melt. Unlike the conventional steel-making processes, this is not an autogenous process and requires an external heat source carefully designed to maintain the iron melt in its molten state through out the process. Here, we propose that the CO released from the BOF or from the present OSRAM converter is to be recovered and burnt in a combustion chamber with stoichoimetric quantities of Oxygen in a closed system. The output of this chamber shall consist of hot gases comprising of $CO_2$ and small quantities of CO. It is desirable to maintain the partial pressure of $CO_2$ close to 1 atm. $CO_2$ being unstable at temperatures above $1100°K$ shall react with carbon in the melt in order to produce the stable CO. Basically this is the Boudouard reaction which is endothermic in nature. The reaction inside the converter is as follows:

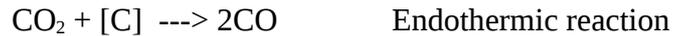

$$CO_2 + [C] \longrightarrow 2CO \qquad \text{Endothermic reaction}$$

The heating mechanism has to ensure that constant temperature is maintained inside the converter at above $1500°\,C$ to ensure the contents remain in a molten state. The other reactions involving oxidation of Mn, Si, Cr etc are not highly favoured as $CO_2$ almost provides an inert atmosphere to these metals. The outgoing hot CO gases are let into the combustion chamber in which it is burnt with pure $O_2$ in order to produce hot $CO_2$ gases that can be recycled into the converter or **a SERIES of** OSRAM converters. Some amount of heat has be recovered to be utilised to generate steam for power generation.

## 4. Discussion:

In order to make $CO_2$ steel-making viable and cost-effective one needs to design an appropriate external heat source to maintain the melt in a liquid state. Jacob Hunt et. al. [10] have shown that microwave irradiation of Carbon dramatically effects the Boudard reaction and the enthalpy of the reaction drops from 183.3 kJ/mol to 33.4 kJ/mol of Carbon at $1100°\,K$. This would lead to substantial energy savings for the process. Hara et.al [11] have attempted pig iron making by micro-wave heating and have made some progress. We are of the opinion that a combination of induction and micro-wave techniques in the design of the converter with the advances in technologies of these two fields might make this $CO_2$ process a success.

To counter this disadvantage, the following are the advantages of the proposed OSRAM-$CO_2$ steel-making process:

1. The process avoids the usage of large amounts of expensive and high energy consuming oxygen.
2. It not only utilises $CO_2$ as the decarburising agent but also recycles and reuses it thereby decreasing the net emissions into the atmosphere thus protecting our dear Earth from greenhouse gases and global warming.
3. In conventional steel-making processes, costly metals such as Mn, Si & Cr are oxidised due to the pure $O_2$ available. In order to avoid this, inert gases such as $N_2$, $H_2$, Argon etc are introduced. In the proposed process, the $CO_2$ itself acts as an inert gas to restrict the oxidation of these costly elements as well as avoids presence of soluble/residual oxygen in the final melt. Generally, so as to remove the residual oxygen in the melt, deoxidising agents like Al, Si, Mn, Ca are utilised as metals or alloys in secondary steel-making process. The quantities of these deoxidising and alloying agents shall be considerably reduced in the current process.
4. As the conventional processes are highly exothermic in nature, the refractory consumption is very large whereas in this process the heat can be closely monitored and controlled by the induction heater along with micro-wave heating and thus protects the refractory lining of the converter.
5. The cost of setting up new steel-making units along with the existing BOF would be

justified in the light of conservation of our dear Earth. Further, the cost-benefit analysis should take into account
a) Additional heat sources to be provided
b) Cost of Oxygen
c) Cost of refractory consumption
d) Cost of deoxidation and alloying agents in the secondary processes
e) Cost of residual/soluble oxygen in the melt that causes defects while casting and shaping of molten steel that effects the quality of steel.

## 5. Conclusions:

We have proposed a new steel-making process named as OSRAM- $CO_2$ Steel-Making process which utilises reuse and recycling of $CO_2$ as the decarburising agent which would help in protecting the environment from global warming and at the same time reducing the use of expensive and energy consuming oxygen required for conventional steel-making. This process shall revolutionise the steel-making all over the world and opens up a whole new area of R&D for working out all the aspects from fundamental calculations for heat & material balances and theormodynamics to highly advanced research analysis of design and optimisation algorithms. We are in the process of designing the pilot studies that need to be taken up to set up a lab-scale furnace to test the proposal of OSRAM- $CO_2$ Steel-Making.

## References:

1. Steel's contribution to a low carbon future, Retrieved from http://www.worldsteel.org/publications/position-papers/Steel-s-contribution-to-a-low-carbon-future.html
2. "The Making, Shaping and Treating of Steel, 10th Edition", United States Steel Co., Chapter 1, Section 7, Modern Steelmaking Processes, pg 24-35
3. John stubbles, "The Basic Oxygen Steel-Making (BOS) Process", Retrieved from https://www.steel.org/Making%20Steel/How%20Its%20Made/Processes/Processes%20Info/The%20Basic%20Oxygen%20Steelmaking%20Process.aspx.
4. C.P. Manning and R.J. Fruehan, "Emerging Technologies for Iron and Steelmaking", JOM, Journal of Electronic Materials and Mettalurgical and Materials Transactions, 53(10)(2001), pp.20-23. International Energy Agency. *Energy Technology Perspective* 179–180 (OECD/IEA, Paris, 2010); available at http://www.iea.org/publications/freepublications/publication/etp2010.pdf
5. Electric Arc Furnaces, https://en.wikipedia.org/wiki/Electric_arc_furnace.
6. Allanore, Antoine, Lan Yin, and Donald R. Sadoway. "A new anode material for oxygen evolution in molten oxide electrolysis." *Nature* 497.7449 (2013): 353-356.
7. P.J. Bhonde, A.M. Ghodgaonkar and R.D. Angal, "Various Techniques To Produce Low Carbon Ferrochrome", Innovation in Ferro Alloy Industry Conference INFACON-XI, 2007, pp. 85-90.
8. Bhonde.P.J. and Angal. R.D. "Solid-state Decarburisation of high-carbon Ferromanganese", Proceedings of the 6th International Ferroalloys Congress, INFACON 6, Volume I, Johannesburg, SAIMM, 1992, pp.161-165.
9. Wang Haijuan, Lidong Teng, and Seshadri Seetharaman. "Investigation of the Oxidation Kinetics of Fe-Cr and Fe-Cr-C Melts under Controlled Oxygen Partial Pressures." *Metallurgical and Materials Transactions B* 43.6 (2012): 1476-1487.
10. Hunt, Jacob, et al. "Microwave-specific enhancement of the carbon–carbon dioxide (Boudouard) reaction." *The Journal of Physical Chemistry C* 117.51 (2013): 26871-26880.
11. Hara, Kyosuke, et al. "Continuous pig iron making by microwave heating with 12.5 kW at 2.45 GHz." *J. Microw. Power Electromagn. Energy* 45.3 (2011): 137-147.